\documentclass{article}
\usepackage{amsmath,graphicx,mlspconf}
\usepackage{comment}
\usepackage{amssymb}
\usepackage{color}

\copyrightnotice{U.S.\ Government work not protected by U.S.\ copyright}

\copyrightnotice{978-1-7281-0824-7/19/\$31.00 {\copyright}2019 Crown}

\copyrightnotice{978-1-7281-0824-7/19/\$31.00 {\copyright}2019 European Union}

\copyrightnotice{978-1-7281-0824-7/19/\$31.00 {\copyright}2019 IEEE}

\toappear{2019 IEEE International Workshop on Machine Learning for Signal Processing, Oct.\ 13--16, 2019, Pittsburgh, PA, USA}

\begin{document}
\title{EEG Signal Dimensionality Reduction and Classification using Tensor Decomposition and Deep Convolutional Neural Networks}
%
\name{Mojtaba Taherisadr, Mohsen Joneidi, and Nazanin Rahnavard\thanks{This material is based upon work supported by the National Science Foundation under Grant No. CCF-1718195.}}
\address{Department of Electrical \& Computer Engineering\\
University of Central Florida, Orlando, USA\\ 
Emails: \{taherisadr@knights, joneidi@ece, nazanin@ece\}.ucf.edu}
%
%
	
%

%


\maketitle
\vspace{-4mm}
\begin{abstract}
\vspace{-5pt}
A new deep learning-based electroencephalography (EEG) signal analysis framework is proposed. While deep neural networks, specifically convolutional neural networks (CNNs), have gained remarkable attention recently, they still suffer from high dimensionality of the training data. 
Two-dimensional input images of CNNs are more vulnerable to be redundant versus one-dimensional input time-series of conventional neural networks.
In this study, we propose a new dimensionality reduction framework for reducing the dimension of CNN inputs based on the tensor decomposition of the time-frequency representation of EEG signals. The proposed tensor decomposition-based dimensionality reduction algorithm transforms a large set of slices of the input tensor to a concise set of slices which are called \emph{super-slices}. Employing super-slices not only handles the artifacts and redundancies of the EEG data but also reduces the dimension of the CNNs training inputs. We also consider different time-frequency representation methods for EEG image generation and provide a comprehensive comparison among them. We test our proposed framework on HCB-MIT data and as results show our approach outperforms other previous studies.      
\end{abstract}
\begin{keywords}
EEG, Convolutional Neural Networks, Time-frequency, Tensor Data Analysis, Dimensionality Reduction.
\vspace{-5pt}
\end{keywords}
\vspace{-3mm}
\section{Introduction}
\vspace{-5pt}
 Electroencephalography (EEG) as a diagnostic tool has been widely used in a wide variety of applications \cite{Game,Feature}. Acquiring and analyzing EEG signals are challenging. Various algorithms have been developed to efficiently process the EEG data, such as frequency analysis \cite{spectro_intro}, wavelet transform \cite{Wavelet_Intro}, filter banks \cite{emd_Intro}, hidden Markov models \cite{HMM_Intro}, support vector machines \cite{SVM_Intro}, and
artificial neural networks \cite{NN_Intro}. \\
\indent All the stated methods involve extraction of hand-crafted features from EEG signals.  
Such hand-crafted feature extraction techniques are ad hoc, time-consuming and may not give the optimal representation of signals. Moreover, for feature extraction one requires a deep domain knowledge to extract effective features. Moreover, the impact of noise interference and particularly artifacts (e.g, eye blink) on data makes the
task of extracting relevant and robust features very challenging \cite{eye_blink}. Recently, deep learning approaches, especially CNNs, have gained significant attention in the field of EEG signal analysis due to their remarkable performances~\cite{CNN_Intro,CNN1_Intro}. CNNs handle the ad-hoc feature extraction process. They also combine the feature extraction and classification steps together. Although CNNs outperform other EEG signal processing methods, they still suffer from the curse of dimensionality of the input training data. Converting each one-dimensional (1D) EEG vector to a two-dimensional (2D) time-frequency (TF) image increases the dimension of the training data, which in turn, increases the required storage space significantly. The challenge of high dimensionality of the CNN model's training data is still open and has to be addressed to improve CNNs' efficiency in terms of storage space and running time.\\
\indent Tensor decomposition is a powerful tool for analysis of high-dimensional data. The collection of TF representations of EEG channels generates a three-way tensor over \emph{time}, \emph{frequency}, and \emph{channel}. This tensor is able to capture temporal and spectral correlations in addition to dependencies of different channels over its third way \cite{tensor}. EEG signals are very sensitive to noise. However, sensing long time series from a large number of channels facilitates utilization of dimensionality reduction techniques in which the impact of noise is diminished in the low-dimensional representation \cite{tensor_noise}.\\
\indent In this paper, we propose an algorithm based on low-rank decomposition of tensors to reduce the size of TF representations of EEG data. Low-rank assumption is a realistic side information for many scenarios in signal processing and communication systems \cite{joneidi16,tari19,ashk16}. 
Firstly, a set of \emph{super-slices}, which are robust superposition of all slices, is computed. Each slice of the input tensor corresponds to one channel. 
Then, the reduced-dimension super-slices are fed to a CNN in order to find the most efficient features and perform classification automatically. Our contributions in this study cab be summarized as following:\vspace{-3mm} 
\begin{itemize}
    \item Proposing a new framework for reducing the dimensionality of TF representation of EEG data based on the tensor decomposition, and feeding the reduced data to a CNN to increase the model's efficiency and decrease its 
    training complexity.\vspace{-3mm} 
    \item Handling noise, artifacts, and redundancies of EEG signals by tensor decomposition-based dimensionality reduction.\vspace{-3mm}
    \item Providing a comprehensive comparison and evaluation of different TF representation approaches for CNN-based EEG signal analysis.\vspace{-2mm}
\end{itemize}

\begin{figure*}[htbp]
\centerline{\includegraphics[scale=0.33]{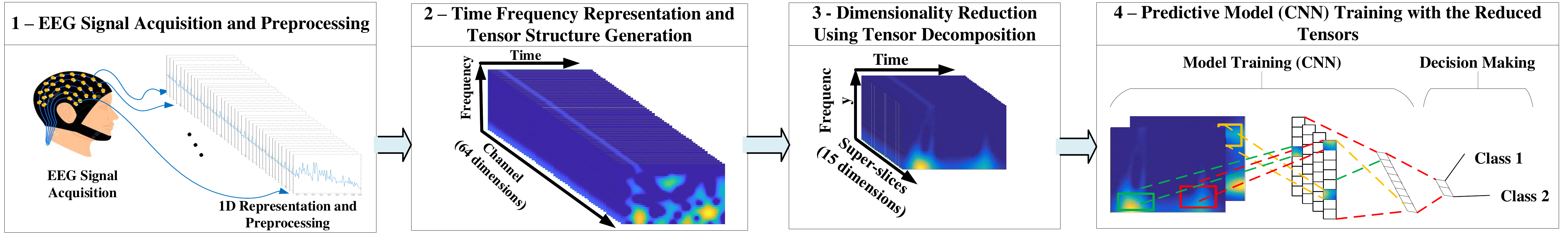}}
\vspace{-3mm}
\caption{\small{Flow chart of the proposed framework. First step visualizes the data acquisition and preprocessing of EEG. In the next step, each segment of the EEG is represented in time-frequency domain as the slices of a 3-way tensor. Finally, tensor decomposition-based technique reduces the tensor to a set of super-slices which is fed to a CNN to train the model and make the decision.}   
}
\vspace{-3mm}
\label{Flow}
\end{figure*}

\textbf{Notations:} Hereafter, vectors, matrices, and tensors are denoted by bold lowercase, bold uppercase, and bold underlined uppercase letters, respectively.
A fiber is defined by fixing every index of a tensor but one.
For example, for $\underline{\boldsymbol{T}} \in \mathbb{R}^{N \times M \times K}$, $\underline{\boldsymbol{T}}_{:,j,k} $ is a vector of length $N$, also known as the mode-1 fiber of $\underline{\boldsymbol{T}}$.  $\boldsymbol{T}_1$, $\boldsymbol{T}_2$, and $\boldsymbol{T}_3$ are unfolded matrices whose columns are fibers of the first, second and third dimensions of $\underline{\boldsymbol{T}}$, respectively.  Slices are two-dimensional
sections of a tensor, defined by fixing all but two indices. 
Moreover, $\circ$ denotes the outer product. The n-mode product of a tensor $\boldsymbol{\underline{X}}$ with a proper sized transformation matrix $\boldsymbol{U}$ is a tensor and denoted by $\boldsymbol{\underline{X}}\times_n \boldsymbol{U}$. It transfers each fiber of the $n^{\text{th}}$ mode of tensor to the corresponding fiber in the final tensor. Mathematically,
$
\boldsymbol{\underline{Y}}=\boldsymbol{\underline{X}}\times_n \boldsymbol{U} \leftrightarrow \boldsymbol{Y}_n=\boldsymbol{U}\boldsymbol{X}_n, 
$
in which $\boldsymbol{X}_n$ and $\boldsymbol{Y}_n$  are unfolded replicas of tensor $\boldsymbol{\underline{X}}$ and $\boldsymbol{\underline{Y}}$ w.r.t. different dimensions. If the vector, $\boldsymbol{u}$ is used instead of the transfer matrix, the result of the n-mode product will be a matrix which is called cotradication of tensor $\underline{\boldsymbol{X}}$ w.r.t. vector $\boldsymbol{u}$.
\vspace{-2mm}
\section{Tensor-based time-frequency dimensionality reduction of EEG signals}
\vspace{-2mm}
In this section, we explain the steps of our proposed framework, as depicted in Fig. \ref{Flow}.
Popularity of CNN has recently increased due to the fact that they outperform classic machine learning approaches. CNN requires 2D images as its input. For this purpose, EEG signals are segmented to equal chunks to then be converted to images using TF representation methods. Each TF method affects the overall performance of the system differently. Therefore, we consider different state-of-the-art TF algorithms to not only optimize performance of our system, but also provide a comprehensive comparison on TF representations of EEG signals. On one hand, more training TF images improve the performance of the CNN models, but on the other hand, it adversely adds to the complexity of the computation. Hence, to reduce the dimensionality of the generated TF images, we employ the \emph{tensor decomposition} technique. Collecting TF representation of EEG segments over $K$ channels, we generate a 3-way tensor over time, frequency, and channel. Tensor decomposition is capable of alleviating artifacts' effects and additionally is able to capture spectro-temporal correlations and dependencies of different channels of EEG signals on its third way. Therefore, as tensor is able to handle artifacts and redundancies of EEG data, we reduce the dimension of the decomposed tensor in its third way which is associated with EEG channels.
After reducing the third dimension of the tensor
to $R$ ($R<<K$), we feed it to CNN to train the model for further predictions. Each step of our proposed algorithm is elaborated in the following.
\vspace{-5mm}
\subsection{Time-frequency representation}\label{tfd}
\vspace{-1mm}
Time-frequency (TF) analysis of an EEG signal is calculating the spectrum at regular time intervals to identify 
the time at which different frequency components present. 
 TF is a suitable representation for non-stationary and multi-component EEG signals because of its ability to describe the energy distribution of the signals over time and frequency simultaneously.  
Previous studies have applied a large number of TF approaches to select a proper methodology for their application, helping to improve the resolution, robustness, precision, or performance. Based on the previous studies, the suitability of a TF approach is data- and application-oriented \cite{Victor}. A review of the recent methods for TF representation reveals that they can be categorized in six groups as follows: Gaussian kernel (GK), Wigner--Ville (WV), spectrogram (SPEC), modified-B (MB), smoothed-WV (SWV), and separable kernel (SPEK).
Reduced interference approaches such as Smoothed-WV are capable of improving the quality of the representation. This is because decreasing the interference results in a reduction in the effect of cross-terms \cite{ava}. Our aim is to assess the mentioned state-of-the-art approaches to determine their performance regarding our specific application in this study (i.e., CNN-based EEG classification).
%
%
\vspace{-8pt}

\subsection{Tensor-based Dimensionality Reduction}
\begin{figure}[b]
\vspace{-2mm}
\centerline{\includegraphics[ height=3.1cm, width=3in]{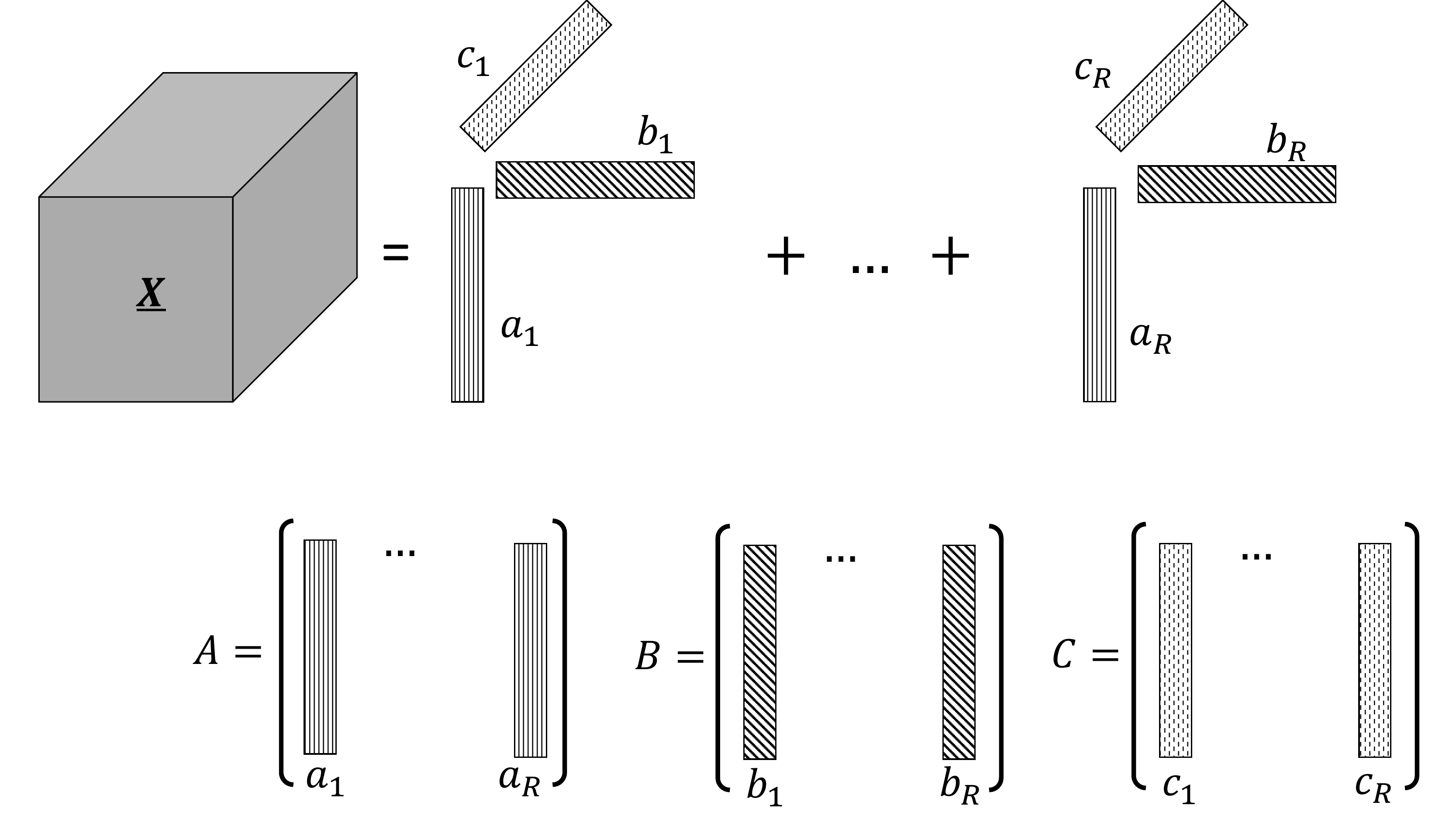}}
\caption{\small{Schematic of decomposition of a rank-$R$ tensor to a summation of $R$ rank-1 tensors.}}
\label{fig:tensor11}
\vspace{-3mm}
\end{figure}

\begin{figure}[h]
\centerline{\includegraphics[width=2.85in]{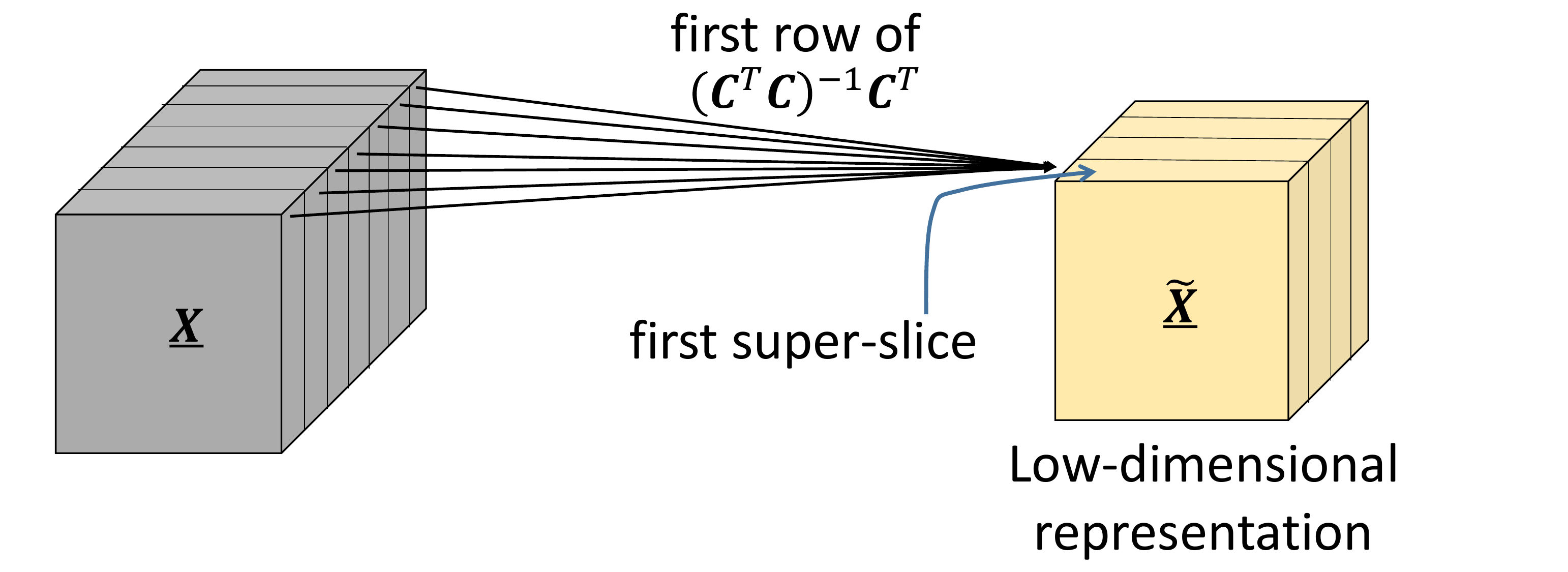}}
\vspace{-3mm}
\caption{\small{The input tensor as a collection of slices is transformed to a set of super-slices. Each super-slice is a superposition of all slices and weights are driven from Matrix $\boldsymbol{P}=(\boldsymbol{C}^T\boldsymbol{C})^{-1}\boldsymbol{C}^T$. For example, the first super-slice is summation of all slices weighted by the first row of $\boldsymbol{P}$.}}
\label{fig:tensor22}
\vspace{-6mm}
\end{figure}

\vspace{-1pt}
As shown in Fig.~\ref{Flow}, the time series of each EEG channel is transformed to a TF representation.
An efficient dimensionality reduction framework is necessary for processing a large set of 2D images generated from 1D EEG data using TF representation.
Let the matrix $\boldsymbol{X}\in \mathbb{R}^{T\times K}$ denote the collection of all time series from $K$ channels and the tensor $\underline{\boldsymbol{X}}\in \mathbb{R}^{T\times F \times K}$ denote the collection of TF representations of the channels. Since time series of different channels are highly correlated, this matrix and the corresponding tensor can be approximated by their low-rank representations. 
In the matrix format, temporal correlation and correlation between channels can be captured via dimensionality reduction techniques such as principle component analysis (PCA). However, for the tensor representation there exist three types of correlation. Efficient dimensionality reduction of tensors implies employing tensor rank decomposition. It should be noted that, performing PCA  on data structured in tensors requires matricization of tensors. After matricization of a tensor, correlation over the unfolded way of the tensor will be neglected. A dimensionality reduction framework that preserves the intrinsic structure of tensors and exploits low-rank tensor decomposition provides a more concise and robust low-dimensional representation. The CANDECOMP/PARAFAC (CP) decomposition of the tensor $\underline{\boldsymbol{X}}$ into $R$ rank-one tensors is given by
\begin{equation}
\label{eq:cp}
    \underline{\boldsymbol{X}}=\sum_{r=1}^R \boldsymbol{a}_r\circ \boldsymbol{b}_r \circ \boldsymbol{c}_r,
\end{equation}
in which, $\boldsymbol{a}_r$, $\boldsymbol{b}_r$ and $\boldsymbol{c}_r$ are called CP factors. Collection of all $\boldsymbol{a}_r$'s in columns of a matrix results in the matrix $\boldsymbol{A}$ and similarly we define $\boldsymbol{B}$ and $\boldsymbol{C}$ matrices. Mode-1 fibers are linear combination of columns of $\boldsymbol{A}$ and similarly mode-2 and mode-3 fibers are linear combination of columns of $\boldsymbol{B}$ and $\boldsymbol{C}$, respectively. The minimum integer $R$ for which~(\ref{eq:cp}) holds is called the rank of $\underline{\boldsymbol{X}}$. Fig. \ref{fig:tensor11} shows the decomposition of a rank-$R$ tensor into a summation of $R$ rank-1 tensors. Definition of rank for tensors is similar to its definition for matrices, however, there are several fundamental differences between matrix rank decomposition (SVD) and tensor rank decomposition (CP) \cite{kolda}.
These fundamental differences encourage us to keep the multi-way structure of the underlying tensor and perform  dimensionality reduction utilizing tensor CP decomposition. Let $\boldsymbol{z}$ denote a mode-3 fiber  of $\underline{\boldsymbol{X}}$. Linear combination of columns of matrix $\boldsymbol{C}$ is able to generate $\boldsymbol{z}$.  
The  representation of any fiber in the third way of $\underline{\boldsymbol{X}}$ in terms of columns of $\boldsymbol{C}$ can be found by solving the 
problem of
   $\Tilde{\boldsymbol{z}}=\underset{\Tilde{\boldsymbol{z}}}{\text{argmin}}\|\boldsymbol{z}-\boldsymbol{C}\Tilde{\boldsymbol{z}}\|_2^2$.
%
 The closed-form solution w.r.t. $\Tilde{\boldsymbol{z}}$ is equal to $(\boldsymbol{C}^T\boldsymbol{C})^{-1}\boldsymbol{C}^T\boldsymbol{z}$. 
Transformation matrix from the original $K$-dimensional space to the reduced $R$-dimension representation is defined by $\boldsymbol{P}=(\boldsymbol{C}^T\boldsymbol{C})^{-1}\boldsymbol{C}^T$. According to this transformation matrix, the original tensor can be reduced as
 $\small{\Tilde{\boldsymbol{\underline{X}}}=\underline{\boldsymbol{X}}\times_3 P}$.
Here, $\Tilde{\boldsymbol{\underline{X}}}$ is the low-dimensional representation of $\boldsymbol{\underline{X}}$ which is a set of super-slices. Mathematically speaking

 $$\underbrace{\Tilde{\underline{\boldsymbol{X}}}_{:,:,r}}_{\;r^{\text{th}} \;\text{super-slice}}=\;\;\underline{\boldsymbol{\boldsymbol{X}}}\times_3 \boldsymbol{P}_{r,:}. $$ 
Here, $\boldsymbol{P}_{r,:}$ indicates the $r^{\text{th}}$ row of $\boldsymbol{P}$. Each super-slice is the contradiction of the original tensor w.r.t. the corresponding row of $ \boldsymbol{P}$.  Fig. \ref{fig:tensor22} shows the relation between super-slices and the slices of the given tensor. Each row of matrix $\boldsymbol{P}$ indicates the weights for generating the corresponding super-slice. Please note that we only reduced the dimension of the third way and the first and second dimensions are preserved in order to extract spectro-temporal patterns using CNN. Using this framework, number of EEG channels is reduced from $K$ to only $R$ super slices ($K>>R$).
%
\subsection{Deep Convolutional Neural Networks (DCNN)}
\indent With CNNs we seek a general-purpose tool for brain-signal decoding capable of extracting a comprehensive set of features without the need for expert knowledge. 
Therefore, we developed a fully supervised CNN model for EEG data analysis. The model takes a super-slice of $\Tilde{\boldsymbol{\underline{X}}}$ (an image) and generates a prediction probability
of belonging to different classes (seizure or non-seizure). We train the model using 
labeled super-slices 
to minimize a $SoftMax$ loss function
with respect to network parameters such as weights and biases using a gradient descent method and network parameters are updated using back propagation.
We  used four main building blocks in the CNN model including convolution,
pooling,
rectified linear unit (ReLU), and
fully connected layer.\\
\indent The primary purpose of convolution layer is to extract features from the input image. Convolution layer preserves the spatial relationship between pixels by learning image features using small squares of input data. The convolution layer performs convolution of input with a set of predefined filters.\\
\indent Spatial pooling reduces the dimensionality of each feature map but retains the most important information. It can be of different types such as maximum and average. In case of Max pooling, we define a spatial neighborhood (for example, a $2\times2$ window) and take the largest element from the rectified feature map within that window.
In practice, Max Pooling has been shown to work better \cite{hinton}.\\
\indent The ReLU is a non-linear activation function that introduces the non-linearity when applied to the feature map. ReLU leaves the size of its input unchanged and it only maps the non-negative values to zero. An additional ReLU has been used after every convolution layer. In fully connected layer each neuron in one layer is connected to all neurons in the next layer. As the output from the convolutional and pooling layers represent high-level features of the input image, we utlize  the fully connected layer to use these features for classifying the input image into various classes based on the training dataset \cite{hinton}.
\vspace{-4mm}
\section{Framework Evaluation and Result Analysis}
\begin{figure} [h]
\centerline{\includegraphics[scale=0.4]{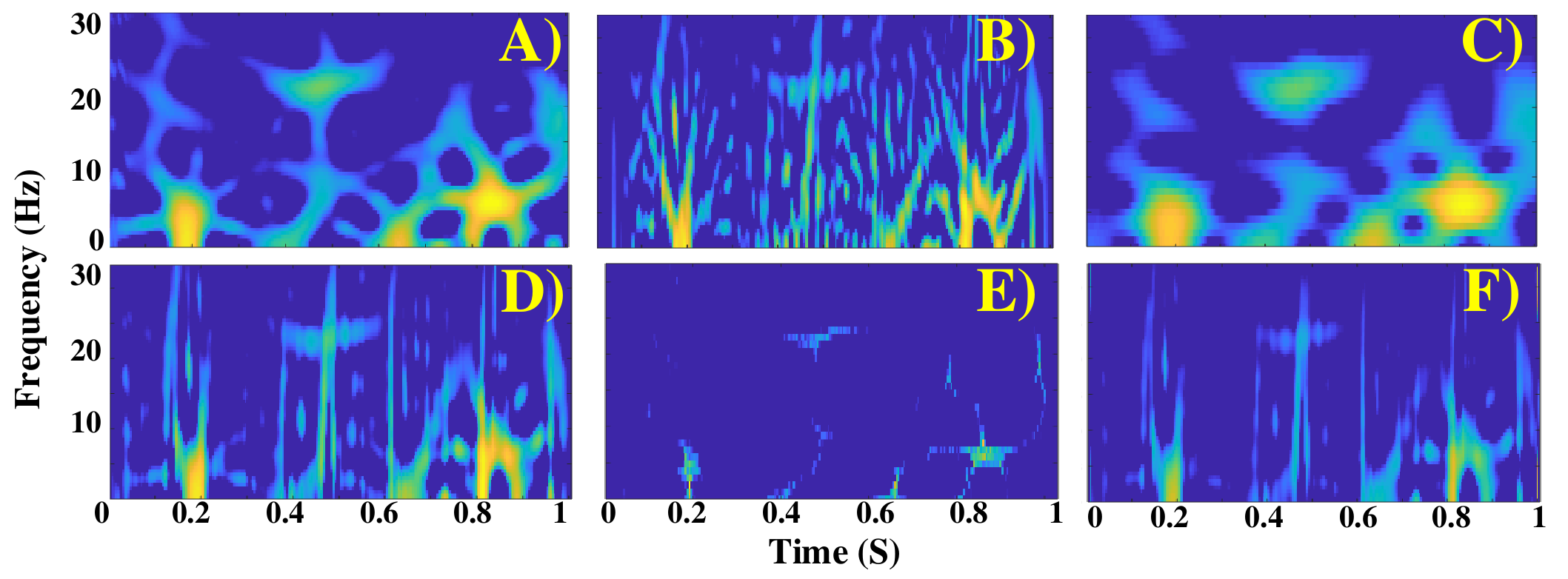}}
\vspace{-3mm}
\caption{TF representations of a 1 second EEG signal using: A) SWV, B) GK, C) WV, D) SPEC, E) MB, and F) SPEK approaches.
}
\vspace{-3mm}
\label{TFs}
\end{figure}

\vspace{-3mm}
We evaluate our proposed method on the  CHB-MIT dataset \cite{shoeb}. 
Different types of epileptic seizures and the diversity of patients contained in this dataset make it ideal for assessing the performance of our framework in realistic settings. In this study, for cross-patient detection,
the goal is to detect whether a 30 second segment of signal
contains a seizure or not, as annotated in the dataset.\\
\indent Different TF methods, 
as discussed in Section \ref{tfd},
 have been considered to generate TF images from EEG segments. Parameters for GKD and MBD have been chosen as $\alpha=0.8$ and $\beta=0.02$, respectively. These values have been selected based on the previous research studies and investigations on theoretical and practical applications of TF representation of EEG signal using GKD and MBD approaches \cite{BoualemB} (Sections 7.4 and 15.5). 
A \emph{Hanning} window is chosen for SPEC and SWVD, with length $Fs/4$ samples, 
where $Fs=256$. Fig. \ref{TFs} illustrates TF representations of a one second interval of EEG signal from one channel using different methods and above-mentioned parameters.

\begin{figure}[h]
\centerline{\includegraphics[scale=0.4]{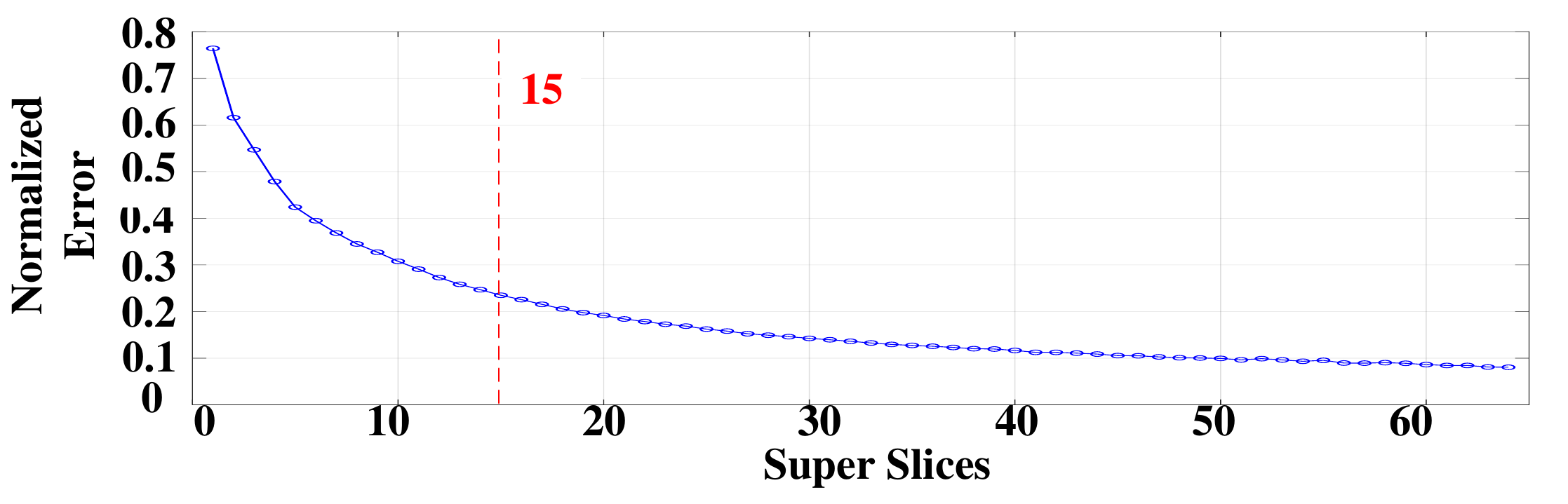}}
\vspace{-4mm}
\caption{\small{Normalized error of CP decomposition versus assumed rank of decomposition.}}
\vspace{-3mm}
\label{fig:tensor_rank}
\end{figure}

\begin{figure}[b]
\vspace{-3mm}
\centerline{\includegraphics[scale=0.26]{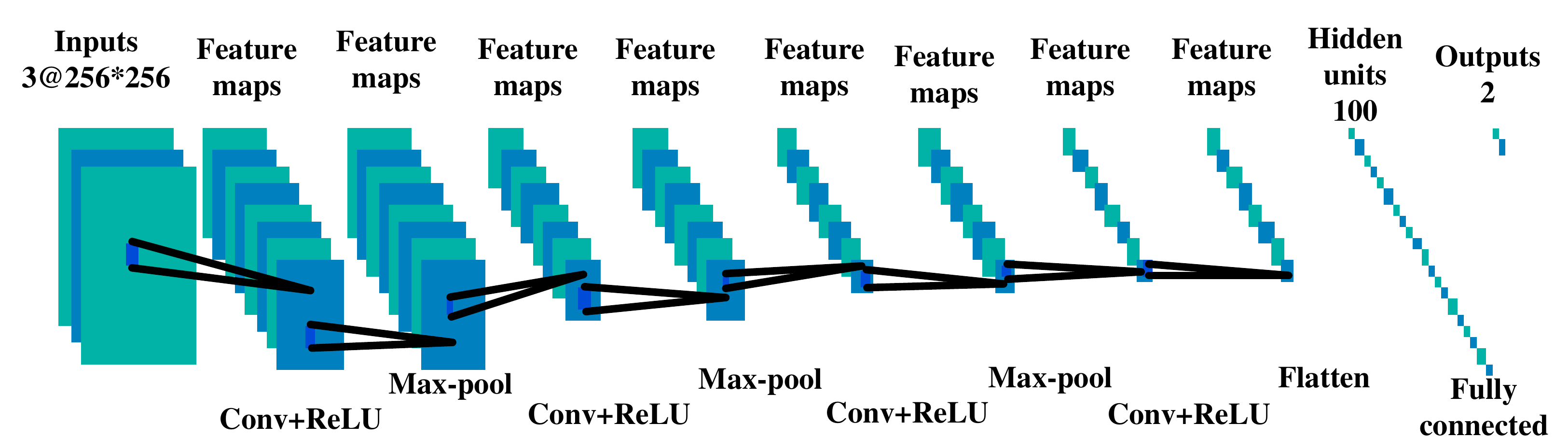}}
\caption{\small{
The CNN architecture proposed in this study. This structure has 10 layers and input image size is 256*256.}} 
\label{CNN_arch}
\vspace{-3mm}
\end{figure}

\indent
Next, Tensor composition has been generated by collecting TF representation of the previous step across all channels. The normalized error of CP decomposition is defined by
\begin{equation}
    {\small \text{normalized error}}=\frac{\|\underline{\boldsymbol{X}}-\sum_r \boldsymbol{a}_r\circ \boldsymbol{b}_r \circ \boldsymbol{c}_r\|_F}{\|\underline{\boldsymbol{X}}\|_F},
    \label{eggg}
\end{equation}
where, $\|.\|_F$ is the Frobenius norm. Fig. \ref{fig:tensor_rank} presents the normalized error of CP decomposition for EEG tensor data. As Fig. \ref{fig:tensor_rank}  demonstrates, increasing the rank of CP decomposition (number of super-slices)  results in a lower normalized error. As Fig. \ref{fig:tensor_rank} shows, the rank around $15$ falls into the interval of normalized error of  $[0.2,\ 0.3]$, which is acceptable for our application.\\  
%
\indent For the DCNN model, the architecture guidelines as mentioned in \cite{changal} were followed. The designed model consists of several layers including (CONV, ReLU, POOL)
 and one fully connected layer as shown in Fig. \ref{CNN_arch}.
Two filter sizes including $2\times2$ and $3\times3$ were tested. 
 ReLU activation layers were used across the CNN after each convolution and pooling pair to bring in element-wise non-linearity.
 In order to estimate the generalization accuracy of the predictive models on the unseen data, 10-fold cross validation (10-CV) was used. 10-CV divides the total input data of $n$ samples into ten equal parts. 
There is no overlap between the test sample set (10\% of data) with the validation and training sample set (90\% of data). The latter set is further divided into 4:1 ratio of training and validation data samples. 
The sets were permuted over 10 iterations to generate an overall estimate of the generalization accuracy. The CNN model was trained using the training set and validation set and tested independently with the testing set. 
Table \ref{modpar} reports the selected parameters to train the CNN model. 
\begin{table}[h]
\centering
\caption{CNN predefined parameters}
\label{modpar}
\footnotesize{
\begin{tabular}{cc}
\hline
\textbf{Parameter} &\textbf{Values}  \\\hline
Learning Rate                       & 0.001  \\
Momentum   Coefficient& 0.9    \\
No. of   Feature Maps                     & 32, 64 \\
No. of Neurons in Fully Connected Layer & 64     \\
Batch   Size    & 40     \\
Epoch Number                         & 19  \\\hline
\end{tabular}}
\vspace{-3mm}
\end{table}

\begin{figure}[t]
\centerline{\includegraphics[scale=0.5]{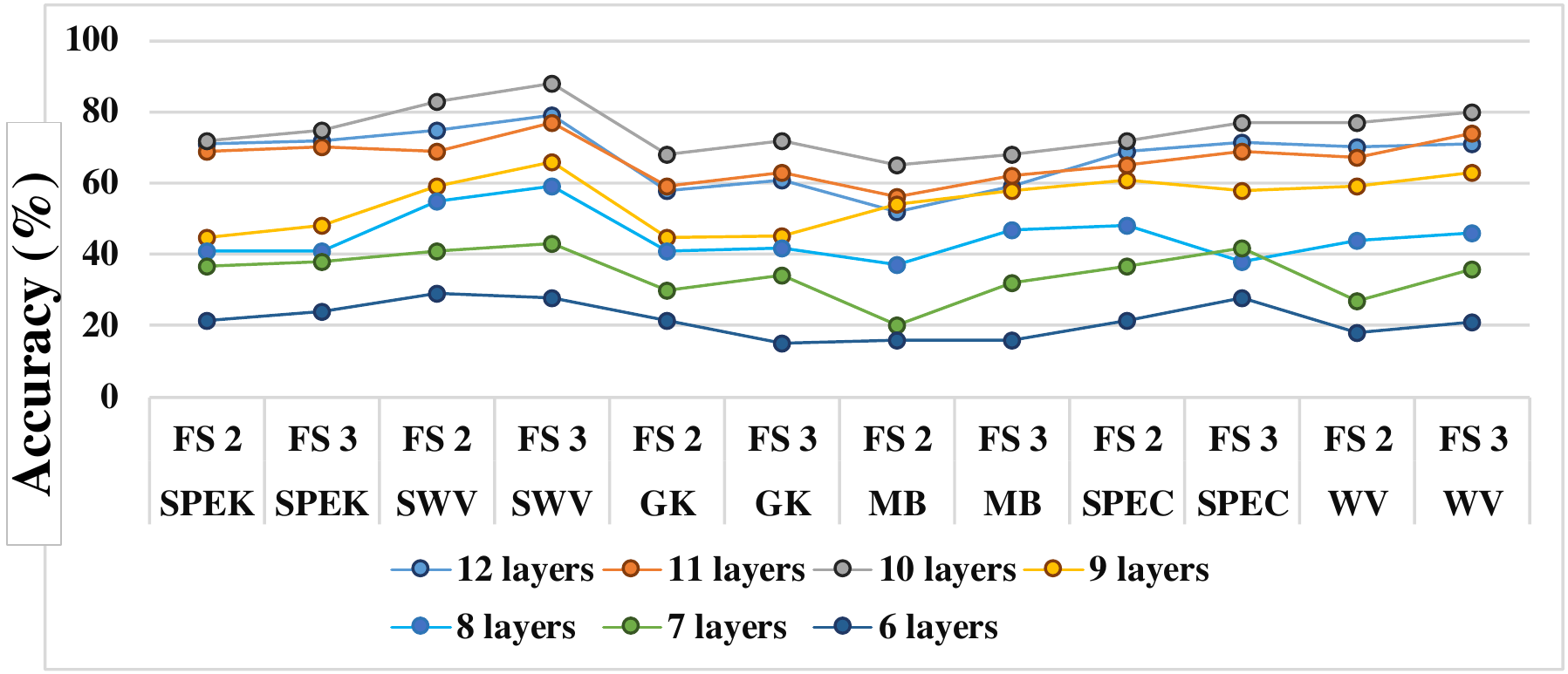}}
\vspace{-3mm}
\caption{\small{Accuracy of EEG signal classification for different TF methods and different CNN parameters. Parameters are different number of layers, and filter sizes are $2\times2$ (FS 2) and $3\times3$ (FS 3). SPEK, SWV, GK, MB, SPEC, and WV indicate different TF representation methods.}}
\vspace{-3mm}
\label{param}
\end{figure}

\indent Then, after defining the parameters of TF representation and CNN model we tested the performance of the designed framework. Fig. \ref{param} depicts the classification accuracy of the proposed framework 
associated with different CNN parameters and TF approaches.
%
$FS$ indicates filter size and SPEK, GK, SWV, WV, MB, and SPEC are TF methods. Seven architectures with different number of layers from 6 to 12, and two filter sizes of $2\times2$ and $3\times3$ are considered. As results present, 10 layers of CNN, filter size of $3\times3$, and SWV TF method outperform other sizes and methods. We use these hereafter. 
\vspace{-3mm}
\subsection{Comparison with Other State-of-the-art and base-line algorithms}
\vspace{-1mm}
In this section we compare our proposed framework with other 1D and 2D baselines. First we consider 1D wavelet transformation as a 1D baseline and then we compare our framework with PCA as a 2D dimensionality reduction baseline. 
\vspace{-6pt}
\subsubsection{Wavelet Transformation}
\vspace{-5pt}
 We extract a set of features from the sub-bands of discrete wavelet transform (DWT). Low- and high-pass filters are repeatedly applied to the signal, followed by decimation by 2, to produce the sub-band tree decomposition to some desired level. 
 DWT of 5 levels was applied to the EEG to reach the approximate frequency ranges of the $\alpha$, $\beta$, $\delta$, and $\theta$ sub-bands \cite{Subasi}.
 After decomposing the signal in each window, features including $average \ power$, 
 $mean$, and  
$standard \ deviation$ of the coefficients were extracted from the sub-bands. 
 Then we feed extracted features to 3 predictive models including complex decision tree (CDT), support vector machine (SVM), and K-nearest neighborhood (KNN). 
The choice of predictive methods was made based on different and complementary properties among them \cite{kuncheva}.
\vspace{-9pt}
\subsubsection{Principal component analysis}
\begin{figure}[t]
\centerline{\includegraphics[scale=0.8]{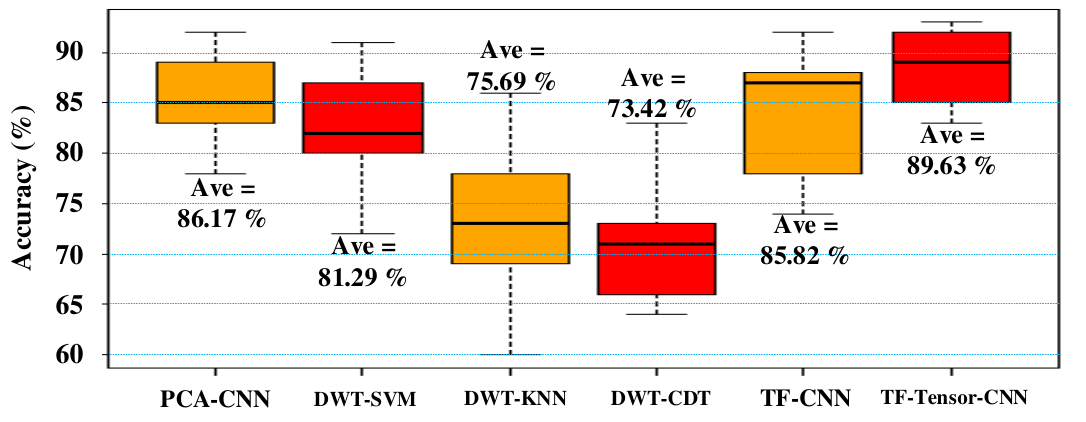}}
\vspace{-3mm}
\caption{\small{Comparison of the classification accuracy of cross-patient seizure detection on CHB-MIT EEG dataset. Each box plot shows 10 iterations of 10 cross validation of the predictive model for the associated method.}}
\vspace{-3mm}
\label{res_all}
\end{figure}

\vspace{-5pt}
We applied PCA to 2D TF data to reduce the dimension and provide the results to compare with our proposed approach. 
We employed PCA to the TF data and analyzed the resulting principal
components (PCs) in order to detect the most descriptive bases of artifacts data. Since the PC space is
orthonormal, we can simply remove the dimensions without affecting others. Based on the results of PCA component contributions, we realized that most of the contribution to the variance of the data ($>85\%$) was summarized
in the first 15 principal components (PCs). Therefore, we kept the first 15 components of the data
for the subsequent predictive model training.\\
\indent Fig. \ref{res_all} summarizes the results of the comparison between 1D and 2D methods considered in this study. It illustrates box plots of 10 iterations of 10-CV algorithm. 
For 1D analysis, as results show, wavelet transform using SVM outperforms others including KNN and CDT. The figure also provides comparison between PCA and the Tensor-based dimensionality reduction schemes and confirms that the tensor-based outperforms the PCA-based dimensionality reduction (callsification accuracy of $89.63\%$ vs. $86.17\%$). Tensor considers all of the channels together and is capable of capturing temporal  and  spectral correlations in  addition to dependencies of different channels over its third way. While PCA works on each TF image separately and it is prone to ignoring the correlations between different channels. Moreover, as Fig. \ref{res_all} shows, the tensor-based dimensionality reduction (TF-tenosr-CNN) framework, due to its capability of reducing the redundancies and handling artifacts, outperforms the TF-CNN framework
 without dimensionality reduction.    \\
\indent Comparing our result (
89\% of accuracy) with previous studies (less than 86\% accuracy), our algorithm has improved the results of cross-patient seizures detection in CHB-MIT dataset \cite{Tzallas,Thodoroff}.

\vspace{-3mm}
\section{Conclusion}
\vspace{-5pt}
In this study, we proposed a new tensor-based framework to enhance the classification accuracy and efficiency of the deep learning models, specifically convolutional neural networks (CNNs), for EEG signals. We proposed a tensor decomposition-based dimentionality reduction of time-frequency (TF) inputs of CNN model to improve its performance in terms of storage space and running time. Our proposed method transforms a large set of slices of the input tensor to a concise set of \emph{super-slices}, which is capable of not only handling the artifacts and redundancies of the EEG data but also reducing the dimension of the CNNs training inputs. We also considered different TF approaches and evaluated their performances to provide a comprehensive comparison of different TF methods for this classification problem. We implemented our proposed method on a publicly available dataset (CHB-MIT). Our results showed the superiority of our scheme compared to the state-of-the-art methods and recent studies. 
\vspace{-9pt}


\begin{small}

\end{small}

\bibliographystyle{IEEEbib}
\bibliography{refs}

\begin{thebibliography}{00}
\vspace{-2mm}

\bibitem{Game}
Taherisadr, M. and Dehzangi, O., 2019. EEG-Based Driver Distraction Detection via Game-Theoretic-Based Channel Selection. In Advances in Body Area Networks I (pp. 93-105). Springer, Cham.
\vspace{-5pt}

\bibitem{Feature}
Dehzangi, O. and Taherisadr, M., 2019. EEG Based Driver Inattention Identification via Feature Profiling and Dimensionality Reduction. In Advances in Body Area Networks I (pp. 107-121). Springer, Cham.
\vspace{-5pt}


\bibitem{spectro_intro}
Tsipouras, M.G., Spectral information of EEG signals with respect to epilepsy classification. EURASIP Journal on Advances in Signal Processing, 2019(1), p.10.
\vspace{-5pt}

\bibitem{Wavelet_Intro}
Taherisadr, Mojtaba, Omid Dehzangi, and Hossein Parsaei. "Single channel EEG artifact identification using two-dimensional multi-resolution analysis." Sensors 17, no. 12 (2017): 2895.
\vspace{-5pt}


\bibitem{emd_Intro}
Gupta, Vipin, Anurag Nishad, and Ram Bilas Pachori. "Focal EEG signal detection based on constant-bandwidth TQWT filter-banks." 2018 IEEE International Conference on Bioinformatics and Biomedicine (BIBM). IEEE, 2018.
\vspace{-5pt}


\bibitem{HMM_Intro}
Mumtaz, Maheen, Mubashira Afzal, and Aleem Mushtaq. "Sensorimotor Cortex EEG signal classification using Hidden Markov Models and Wavelet Decomposition." 2018 IEEE International Symposium on Signal Processing and Information Technology (ISSPIT). IEEE, 2018.
\vspace{-5pt}


\bibitem{SVM_Intro}
Manshouri, Negin, and Temel Kayikcioglu. "A Comprehensive Analysis of 2D \& 3D Video Watching of EEG Signals by Increasing PLSR and SVM Classification Results." arXiv preprint arXiv:1903.05636 (2019).
\vspace{-5pt}


\bibitem{NN_Intro}
Yasmeen, Shaguftha, and Maya V. Karki. "Neural network classification of EEG signal for the detection of seizure." 2017 2nd IEEE International Conference on Recent Trends in Electronics, Information \& Communication Technology (RTEICT). IEEE, 2017.
\vspace{-5pt}

\bibitem{eye_blink}
Dehzangi, O., Melville, A. and Taherisadr, M., 2019. Automatic EEG Blink Detection Using Dynamic Time Warping Score Clustering. In Advances in Body Area Networks I (pp. 49-60). Springer, Cham.
\vspace{-5pt}

\bibitem{CNN_Intro}
Craik, Alexander, Yongtian He, and Jose Luis Pepe Contreras-Vidal. "Deep learning for Electroencephalogram (EEG) classification tasks: A review." Journal of neural engineering (2019).
\vspace{-14pt}

\bibitem{CNN1_Intro}
Yannick, Roy, et al. "Deep learning-based electroencephalography analysis: a systematic review." arXiv preprint arXiv:1901.05498 (2019).
\vspace{-5pt}


\bibitem{tensor}
Cong F, Lin QH, Kuang LD, Gong XF, Astikainen P, Ristaniemi T. Tensor decomposition of EEG signals: a brief review. Journal of neuroscience methods. 2015 Jun 15;248.
\vspace{-5pt}

\bibitem{tensor_noise}
Triantafyllopoulos D, Megalooikonomou V. ''Eye blink artifact removal in EEG using tensor decomposition." InIFIP International Conference on Artificial Intelligence Applications and Innovations 2014 Sep 19 (pp. 155-164). Springer, Berlin.
\vspace{-5pt}






\bibitem{joneidi16}
Joneidi, M., Ahmadi, P., Sadeghi, M. and Rahnavard, N., 2016. ''Union of low-rank subspaces detector". IET Signal Processing, 10(1), pp.55-62.
\vspace{-3pt}
\bibitem{tari19}
Salimitari, M., Joneidi, M. and Chatterjee, M., 2019. AI-enabled Blockchain: An Outlier-aware Consensus Protocol for Blockchain-based IoT Networks. IEEE Global Communications Conference (GLOBECOM) 2019. arXiv preprint is available online, arXiv:1906.08177.
\bibitem{ashk16}
Esmaeili, A., Amini, A. and Marvasti, F., 2016, December. Fast methods for recovering sparse parameters in linear low rank models. In 2016 IEEE Global Conference on Signal and Information Processing (GlobalSIP) (pp. 1403-1407). IEEE.
\vspace{-5pt}

\bibitem{Victor}
Boashash, Boualem, and Victor Sucic. "Resolution measure criteria for the objective assessment of the performance of quadratic time-frequency distributions." IEEE Transactions on Signal Processing 51.5 (2003).
\vspace{-5pt}


\bibitem{ava}
Pedersen, Flemming. "Joint time frequency analysis in digital signal processing." (1997).
\vspace{-5pt}



\bibitem{Boashash.}
Boashash, Boualem. Time-frequency signal analysis and processing: a comprehensive reference. Academic Press, 2015.
\vspace{-4pt}

\bibitem{kolda}
Kolda TG, Bader BW. Tensor decompositions and applications. SIAM review. 2009 Aug 5;51(3):455-500.
\vspace{-4pt}

\bibitem{karpathy2016cs231n}
Karpathy, Andrej. "Cs231n: Convolutional neural networks for visual recognition." Neural networks 1 (2016).
\vspace{-3pt}


\bibitem{zhong2014sensor}
Zhong, Yu, and Yunbin Deng. "Sensor orientation invariant mobile gait biometrics." Biometrics (IJCB), 2014 IEEE International Joint Conference on. IEEE, 2014.
\vspace{-3pt}



\bibitem{BoualemB}
Boashash, Boualem. Time-frequency signal analysis and processing: a comprehensive reference. Academic Press, 2015.
\vspace{-3pt}










\bibitem{changal}
Dehzangi, O., Taherisadr, M. and ChangalVala, R., 2017. IMU-based gait recognition using convolutional neural networks and multi-sensor fusion. Sensors, 17(12), p.2735.
\vspace{-3pt}



\bibitem{Subasi}
Subasi, Abdulhamit. "Automatic recognition of alertness level from EEG by using neural network and Wavelet coefficients." Expert systems with applications 28.4 (2005): 701-711.
\vspace{-3pt}

\bibitem{hinton}
LeCun, Y., Bengio, Y. and Hinton, G., 2015. Deep learning. nature, 521(7553), p.436.
\vspace{-3pt}


\bibitem{shoeb}
Goldberger AL, Amaral LA, Glass L, Hausdorff JM, Ivanov PC, Mark RG, Mietus JE, Moody GB, Peng CK, Stanley HE. PhysioBank, PhysioToolkit, and PhysioNet: components of a new research resource for complex physiologic signals. Circulation. 2000 Jun 13;101(23):e215-20.
\vspace{-3pt}


\bibitem{Tzallas}
Tzallas AT, Tsipouras MG, Tsalikakis DG, Karvounis EC, Astrakas L, Konitsiotis S, Tzaphlidou M. Automated epileptic seizure detection methods: a review study. InEpilepsy-histological, electroencephalographic and psychological aspects 2012 Feb 29. IntechOpen.
\vspace{-3pt}

\bibitem{Thodoroff}
Thodoroff P, Pineau J, Lim A. Learning robust features using deep learning for automatic seizure detection. InMachine learning for healthcare conference 2016 Dec 10 (pp. 178-190).
\vspace{-14pt}








\bibitem{kuncheva}
Kuncheva, Ludmila I. Combining pattern classifiers: methods and algorithms. John Wiley , Sons, 2004.
\vspace{-5pt}


\end{thebibliography}

\end{document}